\documentclass[aps,pra,twocolumn,amsmath,amssymb,showpacs]{
revtex4-1}

\usepackage{graphicx}
\usepackage{color}
\usepackage[T1]{fontenc}
\usepackage{bm}

\usepackage[normalem]{ulem}

\begin{document}

%\preprint{This line only printed with preprint option}

\title{Harmonically trapped Bose-Bose mixtures: a quantum Monte Carlo study
}% Force line breaks with \\
\author{V. Cikojevi\'{c}, L. Vranje\v{s} 
Marki\'{c}}
\affiliation{Faculty of Science, University of Split, Ru\dj era 
Bo\v{s}kovi\'{c}a 33, HR-21000 Split, Croatia}
\author{J. Boronat}
\affiliation{Departament de F\'{\i}sica, 
Universitat Polit\`ecnica de Catalunya, 
Campus Nord B4-B5, E-08034 Barcelona, Spain}

\date{\today}% It is always \today, today,

\begin{abstract}
We study a harmonically confined Bose-Bose mixture using quantum Monte Carlo 
methods. Our results for the density profiles are systematically compared with 
mean-field predictions derived through the Gross-Pitaevskii equation in the 
same conditions. The phase space as a function of the interaction strengths and 
the relation between masses is quite rich. The miscibility criterion for the 
homogeneous system applies rather well to the system, with some discrepancies 
close to the critical line for separation. We observe significant differences 
between the mean-field results and the Monte Carlo ones, that magnify when the 
asymmetry between masses increases. In the analyzed interaction regime, we 
observe universality of our results which extend beyond the applicability regime 
for the Gross-Pitaevskii equation.
\end{abstract}

\maketitle

%\tableofcontents

\section{\label{sec:introduction}Introduction}
Ultracold Bose-Bose mixtures are currently produced in many laboratories 
offering the unique possibility of studying the interplay between two 
Bose-Einstein condensates 
(BEC)~\cite{hall,maddaloni,papp,thal,sugawa,carron,pasquiou,ferrier,wacker,bien,bisset,schultze}.
 The high tunability of the atomic interactions,
thanks to the presence of Feshbach resonances, and the combinations 
between different elements, different isotopes or different hyperfine 
levels enrich dramatically the interest in its physical study. Before the 
adventure of BEC gases, the study of Bose-Bose mixtures was purely academic 
since there was not such a stable system in Nature. Nevertheless, its 
stability was deeply studied in connection with the stable $^3$He-$^4$He 
mixture at low $^3$He concentration. In particular, it was proved that the 
Bose-Bose mixtures of isotopic Helium are always unstable and phase separate 
and that only the right consideration of the Fermi nature of $^3$He atoms could 
account for its finite miscibility~\cite{massey,miller,kurten,tapash}. 

The very low density of the BEC gases makes feasible a theoretical description 
where atomic interactions are modeled by a single parameter, the $s$-wave 
scattering length. The miscibility of bulk Bose-Bose mixtures can be easily 
derived within mean-field theory~\cite{ho,pu,ohberg} and the predictions of 
this approximation 
account well for the observed properties in different experiments. However, 
quite recently this simple argument has been questioned when the mixtures are  
harmonically trapped and a new parameter, based on the shape of the density 
profiles, has been suggested~\cite{lee}. The theoretical descriptions rely on 
the 
Gross-Pitaevskii (GP) equation whose range of applicability has been assumed to 
fit in the relevant experimental setups. Recently, it has been proved that if 
the interspecies interaction is attractive, instead of repulsive, the mixture 
can be stable due to the Lee-Huang-Yang correction which cancels the 
mean-field collapse~\cite{petrov}. This partial cancellation between attractive 
interspecies and repulsive intraspecies interactions can result in a self-bound 
(liquid) system, whose 
existence has been checked theoretically by exact quantum Monte Carlo 
calculations~\cite{viktor} and recent experiments with mixtures of ultracold 
$^{39}$K atoms in different internal states~\cite{cabrera,semeghini}.

In the present paper, we report results of harmonically trapped repulsive 
Bose-Bose mixtures in different interaction regimes and with different masses 
for the constituents. Our approach is microscopic and relies on the use of 
quantum Monte Carlo methods able to solve exactly a given many-body Hamiltonian 
for Bose systems (within some statistical noise). The number of particles of 
our system is much smaller than the typical values used in GP calculations due 
to the complexity of our approach but this allows for an accurate study 
of effects 
going beyond the mean-field treatment. The numerical simulations have been 
carried out 
for different combinations of the interaction strengths, covering the mixed and 
phase separated regimes. In agreement with previous GP results, the miscibility 
rule derived for homogeneous gases fails to describe some of the results.

The rest of the paper is organized as follows. In the next Section we introduce 
the theoretical method used for the study. Sec. III contains the density 
profiles corresponding to the points of the phase diagram here analyzed, both 
in miscible and phase-separated regimes. 
In Sec. IV, we study the scaling in terms of the GP interaction strength. Sec. 
V discusses the universality of our results by changing the model potential.
Finally, Sec. VI reports the main conclusions of our work.

\section{\label{sec:method}Methods}

We study a mixture of two kind of bosons with masses $m_1$ and $m_2$, 
harmonically confined and at zero temperature. The Hamiltonian of the system is
\begin{eqnarray}
H & = &  -\frac{\hbar^2}{2} \sum_{\alpha=1}^{2} \sum_{i=1}^{N_{\alpha}} 
\frac{\nabla_i^2}{m_{\alpha}} +
		\frac{1}{2}\sum_{\alpha, \beta=1}^{2} \sum_{i_{\alpha}, 
j_{\beta=1}}^{N_{\alpha}, N_{\beta}} V^{(\alpha, \beta)} 
(r_{i_{\alpha}j_{\beta}}) \nonumber \\ 
& &		+		\sum_{\alpha=1}^{2} \sum_{i}^{N} V_{\rm 
ext}^{(\alpha)} (\mathbf{r}_{i})
\label{hamiltonian} \ .
\end{eqnarray}
The mixture is composed of $N=N_1+N_2$ particles, with $N_1$ and $N_2$ bosons 
of type 1 and 2, respectively. The interaction between particles is modeled by 
the potentials $V^{(\alpha, \beta)} (r_{i_{\alpha}j_{\beta}})$ and the 
confining potential is a standard harmonic term, with frequencies that can be 
different for each species,
\begin{equation}
V_{\rm ext}^{(\alpha)} (\mathbf{r})
        =   \frac{1}{2} m_{\alpha} \omega_{\alpha}^2 r^2 \ .
\label{harmonic}
\end{equation}

The many-body problem is solved by means of the diffusion Monte Carlo method 
(DMC), a nowadays standard tool for the ab initio study of quantum fluids and 
solids. DMC solves stochastically the imaginary-time $N$-body Schr\"odinger 
equation in an exact way for bosons within some statistical noise~\cite{dmc}. 
In a natural 
way, DMC recovers the mean-field results when the system is dilute enough and 
can go much beyond since no perturbative approximations are assumed. In brief, 
DMC turns the  Schr\"odinger equation into a diffusion process in imaginary 
time with a branching term that replicates the energetically 
favorable configurations and eliminates the rest. In order to reduce the 
variance in the statistical averages it is usual to introduce importance 
sampling through a trial wave 
function, that drives the random walk away from singularities and focuses the 
sampling where one reasonable expects that the ground-state wave function is
large. In the present problem, we have chosen a Jastrow model for the trial 
wave function,
\begin{eqnarray}
\Psi(\mathbf{R}) & = &
    \prod_{1=i<j}^{N_1} f^{(1,1)}(r_{ij})
\prod_{1=i<j}^{N_2}f^{(2,2)}(r_{ij}) \prod_{i,j=1}^{N_1, 
N_2}f^{(1,2)}(r_{ij}) \nonumber \\
            & & \times \prod_{i=1}^{N_1} h^{(1)}(r_i)
		\prod_{i=1}^{N_2} h^{(2)}(r_i) \ ,
\label{trialw}
\end{eqnarray}
with $\mathbf{R}=\{\mathbf{r}_1,\ldots,\mathbf{r}_N\}$, $f^{(\alpha,\beta)}(r)$ 
the two-body Jastrow factors accounting for the pair interactions between 
$\alpha$ and $\beta$ type of atoms, and $h^{(\alpha)}(r)$ the one-body terms 
related to the external harmonic potential.

As one of the objectives of our work is to estimate the validity regime for a 
mean-field approach we have also studied the problem by solving the 
Gross-Pitaevskii (GP) equations. In this case, one assumes contact interactions 
between particles,
\begin{equation}
        V^{(\alpha, \beta)} (r_{i_{\alpha}j_{\beta}})
            =
            g_{\alpha \beta} \, \delta(|\mathbf{r}_{i_{\alpha}} - 
\mathbf{r}_{j_{\beta}}|) \ ,
\label{contact}
\end{equation}
with strengths
\begin{equation}
            g_{\alpha, \beta} = \frac{2 \pi \hbar^2 a_{\alpha 
\beta}}{\mu_{\alpha\beta}} \ ,
\label{gfactors}
\end{equation}
with $\mu_{\alpha\beta}^{-1}=m_{\alpha}^{-1}+m_{\beta}^{-1}$ the reduced mass and $a_{\alpha 
\beta}$ the $s$-wave scattering length of the two-body interaction between 
$\alpha$ and $\beta$ particles.

With the Hartree-Fock \textit{ansatz},
\begin{equation}
            \Psi(\mathbf{R})
            =
            \prod_{i=1}^{N_1}
            \phi_{1}(\mathbf{r}_i,t)
            \prod_{j=1}^{N_2}
            \phi_{2}(\mathbf{r}_j,t)
\label{hartree}            
\end{equation}
one obtains the coupled GP equations for the mixture~\cite{ho},
\begin{eqnarray}
            i \hbar \frac{\partial \phi_{1} (\mathbf{r},t)}{\partial t}
           & = &
            \left(
            -\frac{\hbar^2}{2 m_{1}} \nabla^2
            +
            V_{\rm ext}^{(1)}(\mathbf{r})
            +
            g_{11} |\phi_{1}(\mathbf{r},t)|^2  \right.  \nonumber \\
         & &    +       g_{12} |\phi_{2}(\mathbf{r},t)|^2
            \bigg) 
            \phi_{1} (\mathbf{r},t) \ ,
\label{gp1}
\end{eqnarray}
\begin{eqnarray}
	i \hbar \frac{\partial \phi_{2} (\mathbf{r},t)}{\partial t}
	& = &
	\left(
	-\frac{\hbar^2}{2 m_{2}} \nabla^2
	+
	V_{\rm ext}^{(2)}(\mathbf{r})
	+
	g_{22} |\phi_{2}(\mathbf{r},t)|^2 \right. \nonumber \\
 & &	+
	g_{12} |\phi_{1}(\mathbf{r},t)|^2
	\bigg) 
	\phi_{2} (\mathbf{r},t) \ .
\label{gp2}
\end{eqnarray}
We have solved the GP equations (\ref{gp1},\ref{gp2}) by imaginary-time 
propagation using a 4th order Runge-Kutta method.

\section{\label{sec:phases}Phase space}

We have explored the phase space of the Bose-Bose mixture using the DMC method 
and, in all cases, we have compared DMC with GP results in the 
same conditions. We 
drive our attention to the density profiles of both species since our main goal 
is to determine if the systems are miscible or phase separated. The results 
contained in this Section have been obtained by using hard-core potentials 
between the different particles,
\begin{equation}
	V^{(\alpha\beta)}(r) = 
	\begin{cases}
	\infty  , &  r \leq a_{\alpha \beta}   \\
	0 	  ,&  r > a_{\alpha \beta} \\
\end{cases}
\ ,
\label{hardcore}
\end{equation}
with $a_{\alpha \beta}$ the radius of the hard sphere which 
coincides with 
its $s$-wave scattering length. The Jastrow factor in the trial wave function 
(\ref{trialw}) is chosen as the two-body scattering solution, 
$f^{(\alpha,\beta)}(r)=1-a_{\alpha \beta}/r$, and the one-body term corresponds 
to the exact single-particle ground-state wave function, 
$h^{(\alpha)}(r)=\exp(-r^2/(2a_\alpha^2))$. The length $a_\alpha$ is optimized 
variationally but, even for the strongest interaction regime here studied, its 
value is at most $~10$\% larger than the one of the non-interacting system, 
$l_{\text{ho},\alpha}=\sqrt{\hbar/(m_\alpha \omega_\alpha)}$. 
The one-body terms $h^{(\alpha)}(r)$ confine the particles to move within a 
finite volume and thus open boundary conditions are used. The influence of 
internal parameters of the DMC calculations, such as the number of walkers and the
imaginary-time step, is analyzed and the reported results are converged with 
respect to them. The density profiles that we report are derived using the mixed 
estimation since we have checked that the correction introduced by using pure 
estimators~\cite{pure} is at the same level as the typical statistical noise.

\begin{figure}[]
\centering	
\includegraphics[width=0.85\linewidth]{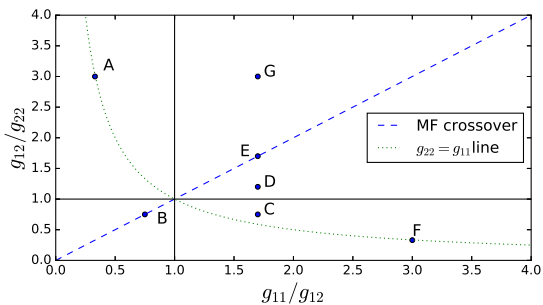}
\caption{Representation of the phase space for the mixture as a function of the 
interaction strengths $g_{\alpha \beta}$. The points correspond to the cases 
studied, with coordinates given in Table 
\ref{table1}. The mean-field theory for 
homogeneous system predicts separation (mixing) for all the points above (below) 
the mean-field critical line (dashed line). The dotted (green) line stands for 
points where $g_{11}=g_{22}$.}
\label{fig1}
\end{figure}

As in Ref.~\cite{lee}, we plot the phase space in terms of the adimensional 
variables 
$g_{12}/g_{22}$ and $g_{11}/g_{12}$. In Fig. \ref{fig1} we plot it showing the 
different regimes, with the line $g_{12}^2=g_{11} g_{22}$ standing for the 
critical line separating miscibility and phase separation, using the mean-field 
criterion.
In the figure, we plot the points which we have studied; they are selected to 
cover the most interesting areas of the phase space. In Table \ref{table1}, we 
report the specific coordinates of the interaction strengths.

\begin{table}[]
\centering
\begin{tabular}{c c c c}
	\hline \hline  \rule{0pt}{11pt}
$\text{Label}$ & $g_{12}/g_{22}$ & $g_{11}/g_{12}$ & $\Delta$     \\ 
                \hline
                $\text{A}$  & $3.0$ & $0.33$ & $-0.89$  \\
                $\text{B}$  & $0.75$ & $0.75$ & 0  \\
                $\text{C}$  & $0.75$ & $1.7$ & $1.27$ \\
                $\text{D}$  & $1.2$ & $1.7$ & $0.42$ \\
                $\text{E}$  & $1.7$ & $1.7$ & 0 \\
                $\text{F}$  & $0.33$ & $3$ & 8 \\
                $\text{G}$  & $3$ & $1.7$ & -0.43 \\
                \hline \hline
		\end{tabular}	
\caption{Representative phase space points analyzed in our study.  
The values of $\Delta$ are obtained from Eq. (\ref{delta}).}
\label{table1}	
\end{table}

As usual in the study of Bose-Bose mixtures we define an adimensional  
parameter $\Delta$,
\begin{equation}
	\Delta = \frac{g_{11} g_{22}}{g_{12}^2} - 1 \ ,
\label{delta}
\end{equation} 	
which classifies the regimes of phase separation ($\Delta < 0$) and miscibility 
($\Delta > 0$) according to the mean-field treatment of bulk mixtures. When 
$\Delta=0$ we are on the critical line separating both regimes (dashed line in 
Fig. \ref{fig1}). In the results reported below, we also calculate the 
parameter $\Delta n$ defined as~\cite{lee}
\begin{equation}
	\Delta n = \frac{\rho_1(0)}{\max \rho_1(r)} - \frac{\rho_2(0)}{\max 
\rho_2(r)} \ ,
\label{deltan}
\end{equation}
which compares the value of the density profiles $\rho_\alpha(r)$ at the 
origin $r=0$ with its maximum value. Then, $\Delta n \simeq 0$ when the peaks 
of both density profiles coincide to be at the origin (mixed state) or when 
the ratio of the central and maximum density is the same for both species, which 
occurs when two species of the same mass separate to two blobs.  
For other types of phase separation $|\Delta n| > 0$.

The density profiles reported in this Section have been obtained with a total 
number of particles $N=200$ and considering a balanced mixture, i.e., 
$N_1=N_2$. We have repeated some calculations considering  
$N_1/N_2=\sqrt{g_{22}/g_{11}}$, which is the optimal balance from the 
mean-field theory~\cite{petrov}, 
and the differences in energy with respect to $N_1=N_2$ are at most 10\%. 
Therefore, we 
concentrate on the balanced mixture. The role of the confining frequencies 
$\omega_\alpha$ is a bit more relevant since we have observed changes in the 
results that can reach the 30\%. In general, the energies are lower when the 
frequencies obey the rule $m_1 \omega_1^2= m_2 \omega_2^2$ which corresponds to 
applying the same harmonic confinement for both species. Therefore, the results 
presented below correspond always to this choice.

\subsection{\label{mixed}$\mathbf{\Delta>0}$}

\begin{figure}[]
\centering
\includegraphics[width=0.9\linewidth]{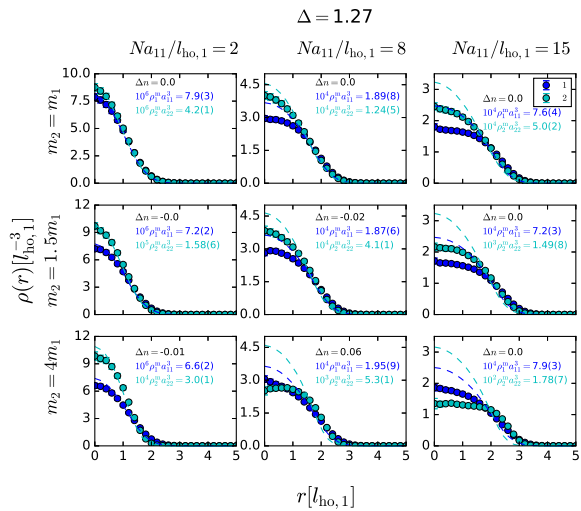}
\caption{Point C of the phase space. The points correspond to the DMC results 
and the lines to the solution of the GP equations for the same system. For each 
case we report the gas parameter of each species calculated at the maximum of 
the density profile.}
\label{pointc}
\end{figure}

When $\Delta>0$ the mean-field criterion predicts
mixing between the 
two species. We have explored the confined system in three different points of 
the phase space. We start with point C, with 
$\Delta=1.27$ and the $g_{\alpha \beta}$ values reported in Table \ref{table1}. 
In Fig. \ref{pointc}, we show the density profiles of both species. In all 
cases, we use as unit length the harmonic oscillator length 
$l_{\text{ho},1}$ of species 1. In the nine subfigures of 
Fig. \ref{pointc}, 
going from left to right we increase the parameter  $N a_{11}/ 
l_{\text{ho},1}$, and from top to bottom we increase the mass of species 2. 
The parameter  $N a_{11}/l_{\text{ho},1}$ is chosen because it is the 
mean-field scaling variable contained in the GP equation. As we keep the total 
number of particles $N$ fixed, increasing that parameter means to increase  the 
scattering length of the 11 interaction and thus making $g_{11}$ larger. As 
the coordinates in the phase space are fixed at the point C, increasing 
$g_{11}$ implies that also the other strengths $g_{12}$ and $g_{22}$ 
increase. Therefore, moving to the right in the panels of Fig. \ref{pointc} 
means an increase of both the interspecies and intraspecies interaction. Moving 
down in the panels, for a fixed value $N a_{11}/l_{\text{ho},1}$, means 
an 
increase of the mass of species 2 and therefore an increase of the scattering 
lengths $a_{12}$ and $a_{22}$ because the couplings $g_{12}$ and $g_{22}$ are 
kept constant.

The density profiles shown in Fig. \ref{pointc} show in all cases a mixed 
state, with $\Delta n \simeq 0$ (\ref{deltan}). In the leftmost column, when 
the interaction is very soft, we appreciate that $\rho^{(\alpha)}(r)$ are 
basically Gaussians, following the shape of the non-interacting gas. When $m_2$ 
grows, they are still Gaussians but slightly different in shape because the 
frequencies are different. The Gaussian profiles disappear progressively moving 
to the right due to the increase of interactions. The comparison between DMC 
and GP shows agreement when the interaction is low and they clearly depart when 
$N a_{11}/l_{\text{ho}}^{(1)}$ grows. The DMC profiles show the emergence of a 
plateau close to $r=0$, in significant contrast with the GP prediction. The 
departure of DMC and GP becomes larger when the gas parameter (included in 
each panel in Fig. \ref{pointc}, calculated near the maximum of each density 
profile), grows.

\begin{figure}[]
\centering
\includegraphics[width=0.9\linewidth]{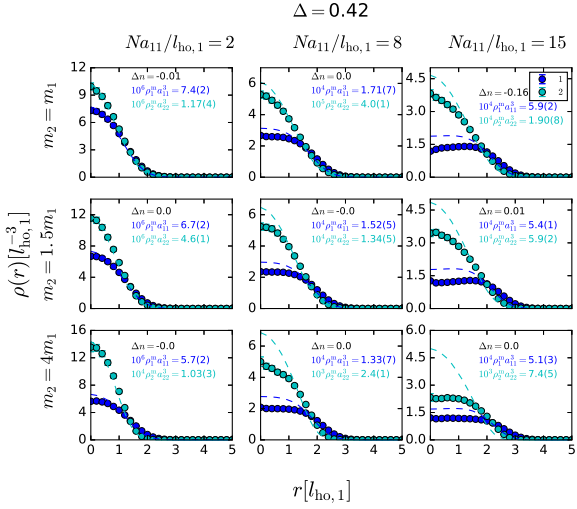}
\caption{Point D of the phase space. The points correspond to the DMC results 
and the lines to the solution of the GP equations for the same system.}
\label{pointd}
\end{figure}

\begin{figure}[]
\centering
\includegraphics[width=0.9\linewidth]{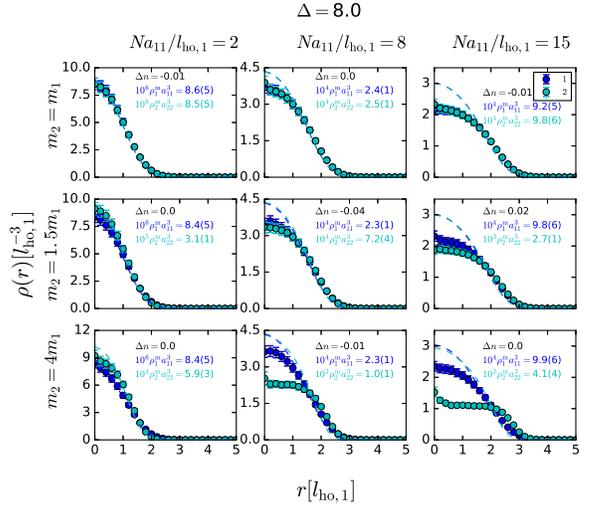}
\caption{Point F of the phase space. The points correspond to the DMC results 
and the lines to the solution of the GP equations for the same system.}
\label{pointf}
\end{figure}

Density profiles for points D ($\Delta= 0.42$) and F ($\Delta=8$) are reported 
in Figs. \ref{pointd} and \ref{pointf}, respectively. The results are 
qualitatively similar to the ones of point C, showing mixing in both cases. 
Point F is deeply located in the mixed part of the phase space (large $\Delta$ 
value) and the agreement with GP is, in this case, quite satisfactory except 
quantitatively  when the mass difference is large and the strength of the 
interaction increases. Point $D$, 
with a small $\Delta$ value, shows slightly more significant  departures 
from GP 
predictions that again increase when the difference in mass between both 
species increases.

\subsection{\label{critical}$\mathbf{\Delta=0}$}

We have studied two points (B and E, see Table \ref{table1}) of the phase 
space which are illustrative of the $\Delta=0$ case. Assuming mean-field theory,
this corresponds to the critical line for mixing in bulk mixtures.

\begin{figure}[]
\centering
\includegraphics[width=0.9\linewidth]{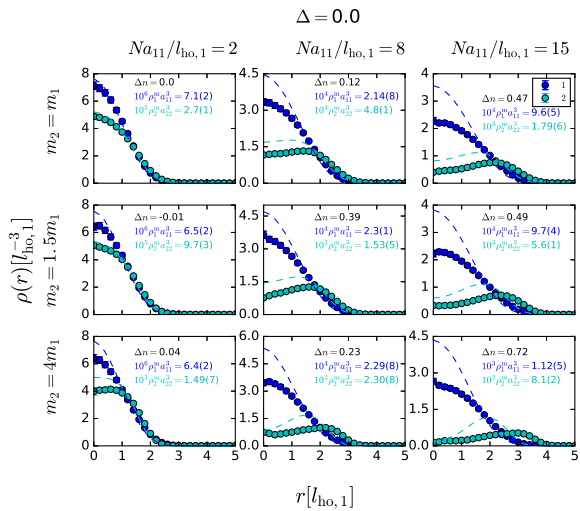}
\caption{Point B of the phase space. The points correspond to the DMC results 
and the lines to the solution of the GP equations for the same system.}
\label{pointb}
\end{figure}

In point B, we have 
$g_{12}/g_{22} = 0.75$ and $g_{11}/g_{12} = 0.75$. With these values,  
$g_{22} > g_{12} > g_{11}$, and thus one expects that species 2 goes out of 
the trap center because it is more repulsive than species 1. This effect is 
emphasized when $m_2>m_1$ because the relation between scattering lengths has 
an additional factor $m_2/m_1$. In Fig. \ref{pointb}, we can see the evolution 
of the density profiles as a function of the interaction strength and mass 
ratio. When the system is only weakly interacting (left column) one appreciates 
Gaussian profiles that coincide with GP predictions. The situation changes when 
the interaction grows (second and third columns) as we can see that the two 
systems start to phase separate, a feature that is measured by the positive 
value of the factor $\Delta n$ (\ref{deltan}). If the difference in mass is 
enlarged, bottom panels, one can see that the phase separation is even more 
clear. In this case, the heaviest component (2) is manifestly going out of the 
center and thus it surrounds the core,  mainly occupied by the species 1. 
Comparison with GP shows that there is a qualitative agreement with DMC but 
quantitatively GP is rather inaccurate, specially when the mass ratio is 
$m_2/m_1=4$.

\begin{figure}[]
\centering
\includegraphics[width=0.9\linewidth]{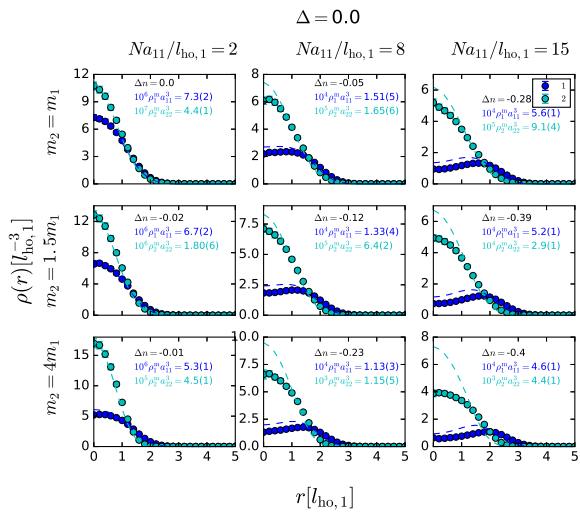}
\caption{Point E of the phase space. The points correspond to the DMC results 
and the lines to the solution of the GP equations for the same system.}
\label{pointe}
\end{figure}

In point E, we are still in the critical line $\Delta=0$ but now the relation 
of interaction strengths is inverted with respect to point B. That is, 
$g_{22} < g_{12}<g_{11}$. Therefore, one now expects  species 2  occupying 
the center and species 1 moving to the external part of the trap. This is 
reflected in the negative values of the parameter $\Delta n$ which are reported 
for every panel in Fig. \ref{pointe}. The increase of the factor $m_2/m_1$ goes 
in reverse direction and slightly compensates the increase in $a_{22}$. 
However, it is shown not to be large enough to change the description. 
Comparing with Fig. \ref{pointb}, the phase separation is not complete because 
one can see that there is always a finite fraction of species 1 close to the 
center.

\subsection{\label{phaseseparated}$\mathbf{\Delta<0}$}

In this subsection, we move to points of the phase space where phase separation 
is expected. We have studied two representative points of the phase space, 
points A and G (see Table \ref{table1}). 

\begin{figure}[]
\centering
\includegraphics[width=0.9\linewidth]{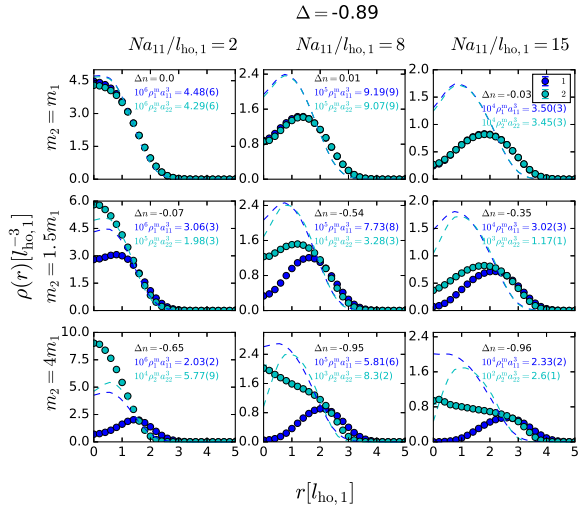}
\caption{Point A of the phase space. The points correspond to 
the DMC results and the lines to the solution of the GP equations for the same 
system.}
\label{pointa}
\end{figure}

\begin{figure}[]
	\centering
	\includegraphics[width=0.6\linewidth, 
angle=270]{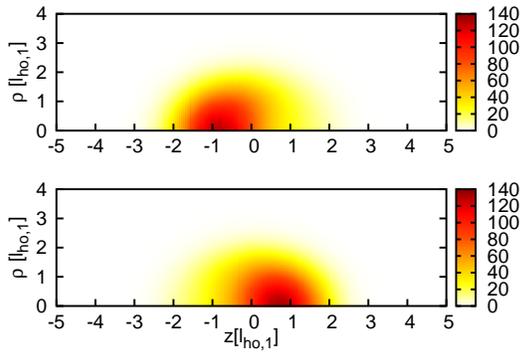}
	\caption{Point A of the phase space for $m_1=m_2$ and $N 
a_{11}/l_{\text{ho},1} = 2$. $z$-axis corresponds to the line going through the 
centers of mass of the two components, while $\rho$ corresponds to the distance 
of a particle from that line. Top and bottom panels stand for the distribution 
of species 1 and 2, respectively.}
	\label{splot1}
\end{figure}

\begin{figure}[]
	\centering
	\includegraphics[width=0.6\linewidth, 
angle=270]{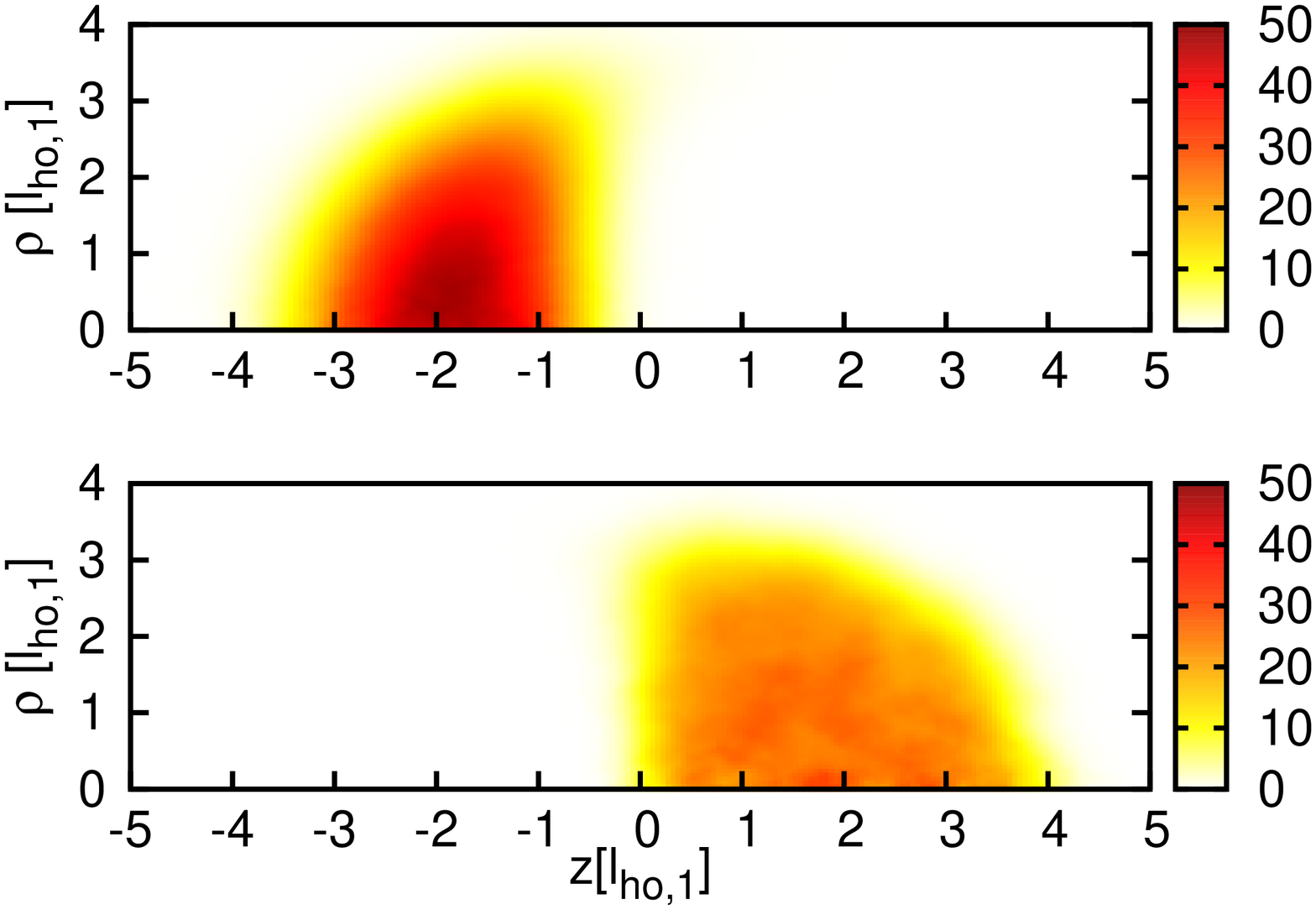}
	\caption{Same as Fig. \ref{splot1} (point A) for $m_2=4m_1$ and $N 
a_{11}/l_{\text{ho},1} = 15$.}
	\label{splot2}
\end{figure}

In point A, $g_{12}/g_{22} = 3$ and $g_{11}/g_{12} = 1/3$ producing 
$\Delta=-0.89$. The relation of strengths is now 
$g_{11}=g_{22} < g_{12}$. By going from left to right in the panels of Fig. 
\ref{pointa} one can see that the mixture phase separates when the interaction 
between atoms is more important than the one-body confining harmonic 
potential. When the masses of both species are equal, one identifies a phase 
separation in form of two symmetric separated blobs~\cite{pu,ohberg}, similar 
to what one 
would observe in 
a bulk system. This is also consistent with $\Delta n = 0$. Again, the 
mean-field prediction becomes quantitatively worse, as the interaction 
strength and the difference in mass between the two components increase. 
However, although it is not visible from the radial profiles, in all cases of 
point A there is at least partial phase separation in two blobs. In order to 
show this, we have calculated the $P(z,\rho)$ distribution, where  the 
$z$-direction is defined as a line passing through the two centers of mass, 
with the second component being in the positive $z$-direction. $z=0$ is the 
geometric center between the two centers of mass. The second variable, $\rho$, 
is just the distance of a single particle from this line. Results are 
normalized such that $\int 2 \pi \rho d\rho dz P(z, \rho) = N/2$. The results in 
the two illustrative cases are presented in Fig. \ref{splot1} and \ref{splot2}. 
In the case of equal masses and $N a_{11}/l_{\text{ho},1} = 2$ (Fig. 
\ref{splot1}) although both species overlap significantly, the maxima of 
their probability distributions are clearly separated. In fact the average 
distance of their centers of mass is about $0.4 l_{\text{ho},1}$. Increasing  
$N a_{11}/l_{\text{ho},1}$ the overlap between the two species decreases and a 
clear two-blob structure becomes visible, with the average distance between the 
two centers of mass becoming $3 l_{\text{ho},1}$. The increase of the mass 
difference of the two species also favors their separation. The extreme case of 
$m_2 = 4m_1$ and  $N a_{11}/l_{\text{ho},1} = 15$ is shown in Fig. \ref{splot2}. 
We observe that the two species are clearly separated in two blobs and more 
spread in both the $\rho$ and $z$ directions than in Fig. \ref{splot1}, due to 
the  larger repulsive interspecies interaction. DMC predicts the more 
massive component to be closer to the center of the trap, unlike GP. $|\Delta 
n|>0$ is quite large, but remarkably has opposite sign in GP and DMC.


\begin{figure}[]
\centering
\includegraphics[width=0.9\linewidth]{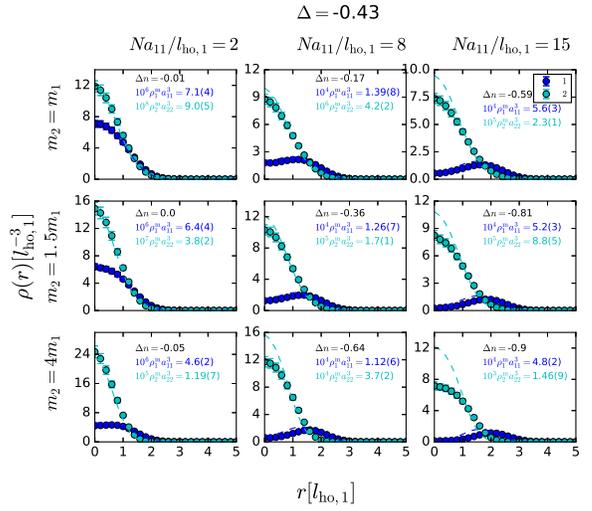}
\caption{Point G of the phase space. The points correspond to 
the DMC results 
and the lines to the solution of the GP equations for the same system.}
\label{pointg}
\end{figure}

\begin{figure}[]
	\centering
	\includegraphics[width=0.6\linewidth, 
	angle=270]{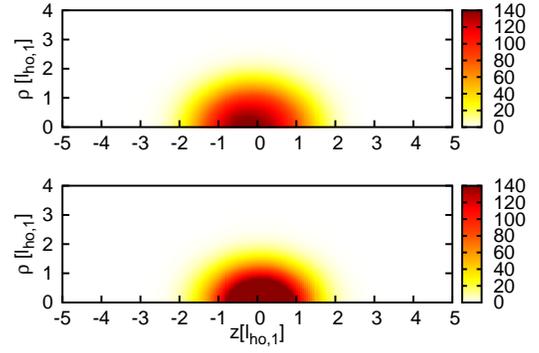}
	\caption{Same as Fig. \ref{splot1} (point G) for $m_2=m_1$ and $N 
		a_{11}/l_{\text{ho},1} = 2$.}
	\label{splot_g1}
\end{figure}

\begin{figure}[]
	\centering
	\includegraphics[width=0.6\linewidth, 
	angle=270]{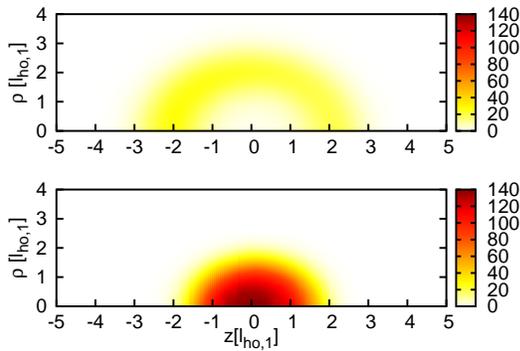}
	\caption{Same as Fig. \ref{splot1} (point G) for $m_2=4m_1$ and $N 
		a_{11}/l_{\text{ho},1} = 15$.}
	\label{splot_g2}
\end{figure}

In the last point (G), one expects phase separation because  
$\Delta=-0.43$. The 
relation of strengths is now $g_{11} > 
g_{12} > g_{22}$. We clearly observe a 
phase separated system for medium and large interactions and a mixed one when 
$Na_{11}/l_{ho,1}=2$. The lighter particle moves progressively out of the 
center and finally, when the relation of masses is large, it surrounds 
completely the heavier one, which occupies the center of the trap. Contrarily 
to point A, here the GP description is in nice agreement with the DMC data even 
when the difference in masses is large. To get a better visualization of this 
point G, we show in Figs. \ref{splot_g1} and \ref{splot_g2} the function  
$P(z,\rho)$ defined above in the same conditions as in Figs. \ref{splot1} and 
\ref{splot2}. We can see that Fig. \ref{splot_g1} shows a mixed configuration 
whereas Fig. \ref{splot_g2} reports a two ring structure, in agreement with 
what is observed in the density profiles of this point (Fig. \ref{pointg}).


\section{\label{scaling}Scaling with the interaction parameter ${\bf N 
a_{11}/l_{\textbf{ho},1}}$ }

In the previous Section, we have plotted the density profiles considering the 
parameter $N a_{11}/l_{ho,1}$ as a scaling parameter to determine the 
strength of 
the interactions. For a given value of this adimensional parameter we have then 
changed the relation of masses between the two species. This parameter has been 
taken from the GP equation for a single Bose gas harmonically confined where 
it is proved to be the \textit{only}  input parameter of the calculations. In 
this section, we check if this is also true in the case of mixtures, both in 
regimes where DMC results and GP ones essentially coincide and in others where 
we have observed significant discrepancies. 

We have chosen two illustrative cases of both situations. In particular, 
points A and E of the phase space (see Table \ref{table1}). In both cases we 
have changed independently $N$ and $a_{11}$ in such a way to keep the GP 
parameter as equal. To this end, we have used a system with $N=100+100$ and a 
smaller one, composed by half the number of particles $N=50+50$.

\begin{figure}[]
\centering
\includegraphics[width=0.9\linewidth]
{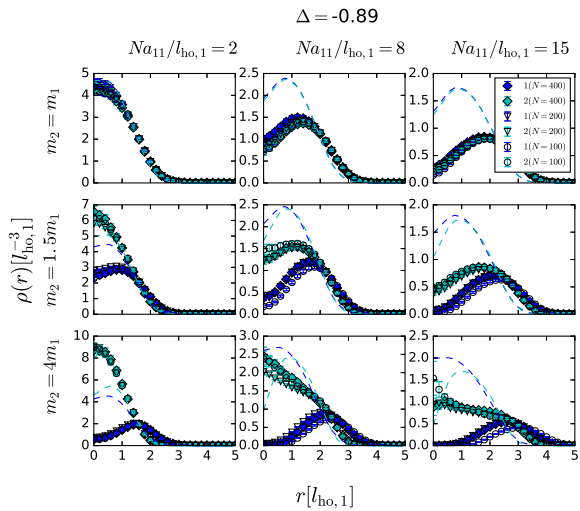} 
\caption{Scaling on the GP parameter in point A of the phase space. 
Different points correspond to different number of particles $N$.
}
\label{pointascal}
\end{figure}

In Fig. \ref{pointascal}, we report the results of this analysis for point A. 
The results of the density profiles are all normalized to sum up to 200 in order 
to make the comparison easier. As we commented in the previous Section, point A 
is the one where we have observed the largest departures from the GP results. 
The figures shows excellent agreement when the interaction is low and some 
discrepancies when the GP parameter grows. However, the effect is not dramatic 
and affects only the heaviest species. It is remarkable that even in 
situations like the ones of Fig. \ref{pointascal},  where GP strongly 
departs from DMC, 
one can still observe a very reasonable scaling with the GP interaction 
parameter. Moreover, we can see that our results converge for a number of 
particles $N \geq 200$ and that these converged results still show 
discrepancies with the GP density profiles.

\begin{figure}[]
\centering
\includegraphics[width=0.9\linewidth]
{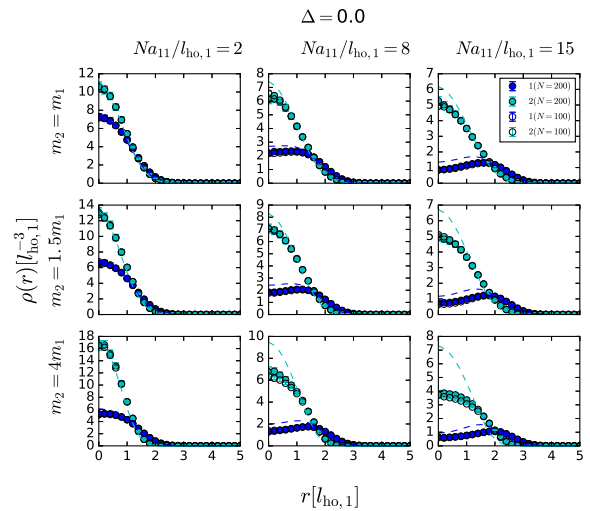} 
\caption{Scaling on the GP parameter in point E of the phase space. Solid and 
open points stand for results with $N=200$ and $N=100$, respectively.
}
\label{pointescal}
\end{figure}

Point E was one of the points where the agreement between GP and DMC was 
better. In Fig. \ref{pointescal}, we study the the dependence of the density 
profile results on the GP parameter as a scaling factor. In this case, the 
agreement is practically perfect because the discrepancies are just of the 
order of the error bars.

\section{\label{universality} Universality test}
A relevant point in our numerical simulations is the influence of the model 
potential on the results. Universality in these terms means that the 
interaction can be fully described by a single parameter, the $s$-wave 
scattering length, as it corresponds to a very dilute system. In all the 
previous results we have used a hard-core model for the interactions between 
the atoms (\ref{hardcore}). In this Section, we compare these results with 
other ones obtained with a 10-6 potential,
 \begin{equation}
 V^{(\alpha\beta)}(r)=\frac{\hbar^2}{2 \mu_{\alpha\beta}}\, V_0 \left[ 
\left(\frac{r_0}{r}\right)^{10}-\left(\frac{r_0}{r}\right)^6 \right] \ ,
\label{lenjones}
 \end{equation}
whose $s$-wave scattering length is analytically known~\cite{pade}. We fix the 
parameter 
$r_0=2 a_{\alpha \beta}$ for all cases and modify the strength $V_0$ to 
reproduce the desired scattering length. In Eq. (\ref{lenjones}), the 
parameter $\mu_{\alpha\beta}$ is the reduced mass, $\mu_{\alpha\beta}=m_\alpha m_\beta/(m_\alpha+m_\beta)$.

Our analysis has been performed for $m_2=1.5 m_1$, $N=100+100$ particles, and 
considering equal harmonic frequencies $\omega_1=\omega_2$. Two interaction 
strengths have been used, $Na_{11}/l_{\text{ho}, 1}=8$ and 15. As in the 
previous 
Section, we have studied points E and A of the phase space, i.e., those 
characteristic of agreement and disagreement with GP.

\begin{figure}[]
	\centering
	\includegraphics[width=0.9\linewidth]
	{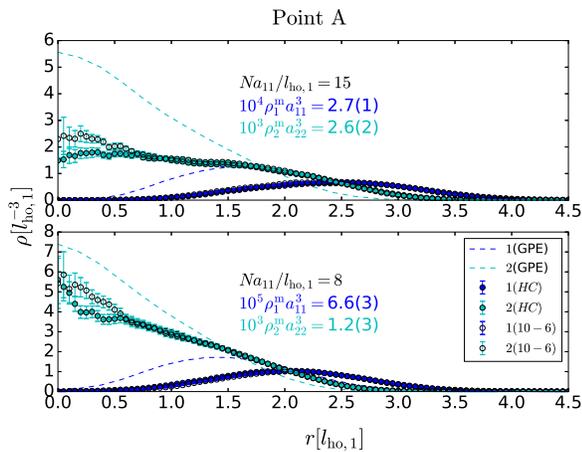} 
	\caption{DMC results for two models are compared to GP results for two values of GP parameter in point A .
	}
	\label{aunivers}
\end{figure}

In Fig. \ref{aunivers}, we show the density profiles of point A, obtained using 
the two model potentials normalized in the same way. We can see an overall 
agreement between both results with only some differences in the estimation of 
the density in the center. Close to $r=0$ the statistical fluctuations are 
bigger due to the normalization in a small volume. This feature is always 
present but we observe that these fluctuations are larger in the case of the 
10-6 potential (\ref{lenjones}). 

\begin{figure}[]
	\centering
	\includegraphics[width=0.9\linewidth]
	{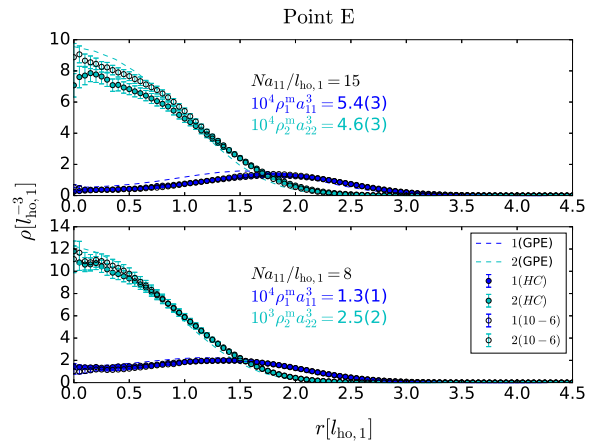} 
	\caption{DMC results for two models are compared to GP results for two values of GP parameter in point E.
	}
	\label{eunivers}
\end{figure}

In Fig. \ref{eunivers}, we analyzed the same for point E in 
which we are closer to an effective mean-field description. The comparison also 
shows a good agreement between results obtained for both potentials, with some 
differences in the inner core of the trap. 

We have verified in other points of the phase space the universality of our 
results and the conclusion is that this is maintained with respect to the 
character of the system (miscible or phase separated) and also the overall 
shape of the density profiles. The influence of the model potential is at the 
scale of our statistical fluctuations, except close to $r=0$ where small 
differences are observed in some cases.

\section{\label{discuss} Discussion}
Using the diffusion Monte Carlo method, able to provide exact results for 
bosonic systems,we have explored the phase space of an harmonically confined 
Bose-Bose mixture at zero temperature. Our results are compared in all cases 
with a mean-field Gross-Pitaevskii calculation in the same conditions. As 
expected, our DMC results agree better with the GP ones when the strength of 
interactions is small and worsens progressively when that grows. In spite of 
the fact that the prediction for miscibility or phase separation is coincident 
in both cases, it is also true that the density profiles can be 
rather different. The relevance of quantum fluctuations, which makes the 
density profiles to depart from the GP ones, depends on the dimensionality of 
the system. It is well known~\cite{astra,edler} that they become much more 
significant in the case of one dimension, since the Lee-Huang-Yang term scales 
as $n^{1/2}$ instead of the three-dimensional law $n^{3/2}$. 

Our results are by construction exact and go beyond mean-field, showing the 
limits of this approach. A systematic trend in which GP fails is the dependence 
on the mass of the two species. When the asymmetry in the masses grows GP 
becomes clearly wrong in some cases. That could be argued to be an effect of 
the interaction model used in DMC but our results contradict this point, within 
the range of interactions here explored.

The tunability of atomic interactions and the possibility of using atoms with 
different mass ratios makes this system specially rich. The phase space shows 
regimes of miscibility and, interestingly, two different situations for phase 
separation, two blobs or in-out spherical separation, depending in the 
relation of masses and interaction strength. It would be very interesting to 
produce in the lab smaller systems, with hundreds of atoms, to check the 
departure of the physics from the mean-field GP treatment. In this way, one can 
start to enter into the realm of fully quantum many-body physics.

\begin{acknowledgments}
This work has been supported in part by the Croatian Science Foundation 
under the project number IP-2014-09-2452 and MINECO (Spain) Grants No. 
FIS2014-56257-C2-1-P and No. FIS2017-84114-C2-1-P. The computational resources of the Isabella cluster at 
Zagreb 	University Computing Center (Srce)  and Croatian National
Grid Infrastructure (CRO NGI) were used.
\end{acknowledgments}

\end{document}